\documentclass[preprint,showpacs,preprintnumbers,amsmath,amssymb]{revtex4}
\usepackage{graphicx}% Include figure files
\usepackage{dcolumn}% Align table columns on decimal point
\usepackage{bm}% bold math

\newcommand{\be}{\begin{equation}}
\newcommand{\en}{\end{equation}}
\newcommand{\bea}{\begin{eqnarray}}
\newcommand{\ena}{\end{eqnarray}}

\begin{document}

%\preprint{GACG/07/2006}

\title{ Warm inflationary  model in loop quantum cosmology }

\author{Ram\'on Herrera}
\email{ramon.herrera@ucv.cl} \affiliation{ Instituto de
F\'{\i}sica, Pontificia Universidad Cat\'{o}lica de
Valpara\'{\i}so,  Avenida Brasil 2950, Casilla 4059,
Valpara\'{\i}so, Chile.}

\date{\today}% It is always \today, today,
             %  but any date may be explicitly specified

\begin{abstract}
A warm inflationary universe model in loop quantum cosmology is
 studied. In general we discuss the condition of inflation  in this
 framework.  By using a chaotic  potential, $V(\phi)\propto \phi^2$, we develop a model
 where   the dissipation  coefficient $\Gamma=\Gamma_0=$ constant.   We  use recent astronomical
 observations  for constraining the parameters appearing in our model.
\end{abstract}

\pacs{98.80.Cq}% PACS, the Physics and Astronomy
                             % Classification Scheme.
%\keywords{Suggested keywords}%Use showkeys class option if keyword
                              %display desired
\maketitle

\section{Introduction}

It is well  know that warm inflation, as opposed to the
conventional cool inflation, presents the attractive feature that
it avoids the reheating period \cite{warm}. In these kind of
models dissipative effects are important during the inflationary
period, so that radiation production occurs concurrently together
with the inflationary expansion. If the radiation field is in a
highly excited state during inflation, and this has a strong
damping effect on the inflaton dynamics, then it is found a strong
regimen of  warm inflation. Also, the dissipating effect arises
from a friction term which describes the processes of the scalar
field dissipating into a thermal bath via its interaction with
other fields. Warm inflation shows how thermal fluctuations during
inflation may play a dominant role in producing the initial
fluctuations  necessary for large-scale structure  formation. In
these kind of models the density fluctuations arise from thermal
rather than quantum fluctuations \cite{62526}. These fluctuations
have their origin in the hot radiation and influence the inflaton
through a friction term in the equation of motion of the inflaton
scalar field \cite{1126}. Among the most attractive features of
these models, warm inflation end when the universe heats up to
become radiation domination; at this epoch the universe stops
inflating and "smoothly" enters in a radiation dominated Big-Bang
phase\cite{Berera:2008ar,jj}. The matter components of the
universe are created by the decay of either the remaining
inflationary field or the dominant radiation field
\cite{taylorberera}.

On the other hand, Loop Quantum Gravity (LQG) is a resulting
nonperturbative background independent approach to quantize
gravity \cite{5}. Here, the geometry in LQG is discrete and the
continuum space-time is obtained from quantum geometry in a large
eigenvalue limit. The application of LQG techniques to homogeneous
space-times results in LQC  which has directed  to important
insights on the resolution of singularities\cite{6,7, 8, 9}.
Within the various conceivable  cosmological models the ones which
are best understood are the Friedmann-Robertson-Walker (FRW)
models \cite{AA}. In this case it has been shown that the quantum
isotropic and homogeneous gravitational degrees of freedom
minimally coupled to the massless scalar field allow non-singular
evolution for the open, closed and flat universes. Here, the
singularity becomes substituted with the smooth Big Bounce. In
this sense the initial  singularity is resolved by the quantum
gravitational repulsion effects. Because of the loop quantum
effect the standard Friedmann equation can be modified by adding a
correction term $\rho^2$ at the scale when $\rho$ becomes
comparable to a critical density $\rho_c\approx 0.82\,G^{-2}$ ($G$
is the Newton's gravitational constant) which is close to the
Planck density. Within the framework of LQC the inflationary model
has been considered in Ref.\cite{good}. Recently, the dynamics of
the interacting dark energy model in Einstein and loop quantum
cosmology was considered in \cite{int1}, and the cosmological
evolution of the  interacting phantom (quintessence) model in loop
quantum gravity was studied in Ref.\cite{int2}.

The main goal of the present work is to investigate the possible
realization of a  warm inflationary universe model, within the
framework of the effective theory of loop quantum cosmology. In
this way, we study warm-LQC model and the cosmological
perturbations, which are expressed in term of different parameters
appearing in our model. These parameters are constrained from the
WMAP 5-year data \cite{WMAP}. Also, we only discuss the  normal
inflation epoch, i.e., after the super-inflation scenario. For a
review of super-inflation, see, e.g., \cite{good,super}.

The outline of the paper is a follows. The next section presents a
short review of the effective theory of LQC. In Section
\ref{secti} we present the warm inflationary phase in this
framework. Section \ref{sectpert} deals with the scalar and tensor
perturbations, respectively.  In Section \ref{exemple} we use a
chaotic potential and $\Gamma=\Gamma_0=$ constant, for obtaining
explicit expression for our model. Finally, Sec.\ref{conclu}
summarizes our findings. We chose units so that $c=\hbar=1$.

\section{Loop quantum cosmology}

The effective Friedmann equation can be obtained by using an
effective Hamiltonian with loop quantum modifications \cite{38,
44, 52} \be
{\cal{H}}_{eff}=-\frac{3}{\kappa\,\gamma^2\,\bar{\mu}^2}\,a\,\sin^2(\bar{\mu}\,\mathfrak{c})+{\cal{H}}_{M},
 \en
where $\kappa=8\pi G$, $\gamma$ is the dimensionless
Barbero-Immirzi parameter ($\gamma\approx 0.2375$ see
Ref.\cite{Bh}), $\bar{\mu}$ is inferred as the kinematical length
of the square loop, and ${\cal{H}}_{M}$ is the matter Hamiltonian.
Here, $\mathfrak{c}$ and $\mathfrak{p}$ are, respectively,
conjugate connection and triad satisfying
$\{\mathfrak{c},\mathfrak{p}\}=\gamma\kappa/3$,  and the relation
with the metric components of the FRW becomes \be
\mathfrak{c}=\gamma\,\dot{a}\,,\;\;\;\mbox{and}\;\;\;\;\;\mathfrak{p}=a^2\,,\label{cp}
 \en
where $a$ represents the scale factor. The modified Friedmann
equation can be found by using Hamilton's equations for
$\dot{\mathfrak{p}}$, \be
\dot{\mathfrak{p}}=\{\mathfrak{p},{\cal{H}}_{eff}\}=\frac{2a}{\gamma\,\bar{\mu}}\,
\sin(\bar{\mu}\,\mathfrak{c})\,\cos(\bar{\mu}\,\mathfrak{c}),\label{dp}
 \en
and from Eq.(\ref{cp}) implies that $\dot{a}$ is given by \be
\dot{a}=\frac{1}{\gamma\,\bar{\mu}}\,\sin(\bar{\mu}\,\mathfrak{c})\,\cos(\bar{\mu}\,\mathfrak{c}).\label{da}
\en Furthermore, the vanishing of the Hamiltonian constraint
implies\cite{38, 44, 52} \be \sin^2(\bar{\mu}\,\mathfrak{c})
=\frac{\kappa\,\gamma^2\,\bar{\mu}^2}{3\,a}\;{\cal{H}}_{M}.\label{sin}\en

From Eqs.(\ref{da}) and (\ref{sin}), the effective Friedmann
equation becomes \be
 \label{newfried}
 H^2=\frac{\kappa}{3}\,\rho\,\left[1-\frac{\rho}{\rho_{c}}\right],
 \en
where  $H=\dot{a}/a$ is the Hubble parameter,
$\rho_{c}=\sqrt{3}\,\rho_{p}/(16\pi^2\gamma^3)$ is the critical
loop quantum density and $\rho_{p}$ is the Planck density equal to
$\rho_{p}=G^{-2}$.

In the following  we will consider a total energy density
$\rho=\rho_\phi+\rho_\gamma$ where $\phi$ corresponds to a
self-interacting scalar field with  energy density, $\rho_\phi$,
given by $\rho_\phi=\frac{1}{2}\dot{\phi}^2+V(\phi)$, and
$\rho_\gamma$ represents  the radiation energy density.

\section{Warm-LQC Inflationary phase \label{secti}}

 The dynamics of the
cosmological model in the warm-LQC inflationary scenario is
described by the equations
 \be \ddot{\phi}+\,3H \;
\dot{\phi}+V'=-\Gamma\;\;\dot{\phi}, \label{key_01}
 \en
and \be \dot{\rho}_\gamma+4H\rho_\gamma=\Gamma\dot{\phi}^2
.\label{3}\en Here $\Gamma$ is the dissipation coefficient and it
is responsible for the decay of the scalar field into radiation
during the inflationary era. In general, $\Gamma$ can be assumed
to be a constant or a function of the scalar field $\phi$, or the
temperature of the thermal bath $T_r$, or
both\cite{warm,Moss,Zhang:2009ge,delCampo:2009xi}. On the other
hand, $\Gamma$ must satisfy $\Gamma>0$  by the Second Law of
Thermodynamics.

Statistical mechanics of quantum open systems has shown that the
interaction of quantum field with a thermal bath can be
characterized by a fluctuation dissipation relation\cite{new0}.
These effects support the idea of introducing a friction term into
the field equation of motion. Here, the friction term
$\Gamma\dot{\phi}$ describes the interaction between the scalar
field $\phi$ and the heat bath. The possibility of warm inflation
arising in realistic particles models has been enhanced by the
decay mechanism in supersymmetric theories, where the inflaton
decays into radiation fields  as a consequence of the heavy
particle intermediate\cite{new}. If the coupling constants are
sufficiently large, these models can lead to warm
inflation\cite{new2}. Dots mean derivatives with respect to time
and $V'=\partial V(\phi)/\partial\phi$.

During the inflationary epoch the energy density associated to the
scalar field is of the order of the potential, i.e. $\rho_\phi\sim
V$, and dominates over the energy density associated to the
radiation field, i.e. $\rho_\phi>\rho_\gamma$.  Assuming the set
of slow-roll conditions, i.e. $\dot{\phi}^2 \ll V(\phi)$, and
$\ddot{\phi}\ll (3H+\Gamma)\dot{\phi}$ \cite{warm}, the Friedmann
equation (\ref{newfried})  reduces  to
\begin{eqnarray}
H^2\approx\frac{\kappa}{3}\,V\left[1-\frac{V}{\rho_c}\right],\label{inf2}
\end{eqnarray}
and  Eq. (\ref{key_01}) becomes
\begin{equation}
3H\left[\,1+R \;\right ] \dot{\phi}\approx-V', \label{inf3}
\end{equation}
where $R$ is the rate defined as
\begin{equation}
 R=\frac{\Gamma}{3H }.\label{rG}
\end{equation}
For the strong  (weak) dissipation  regime, we have $R\gg 1$ ($R<
1$).

We also consider that  during  warm inflation the radiation
production is quasi-stable, i.e. $\dot{\rho}_\gamma\ll 4
H\rho_\gamma$ and $ \dot{\rho}_\gamma\ll\Gamma\dot{\phi}^2$.  From
Eq.(\ref{3}) we obtained that the energy density of the radiation
field becomes
 \begin{equation}
\rho_\gamma=\frac{\Gamma\dot{\phi}^2}{4H},\label{rh}
\end{equation}
which  could be written as $\rho_\gamma= \sigma T_r^4$, where
$\sigma$ is the Stefan-Boltzmann constant and $T_r$ is the
temperature of the thermal bath. By using Eqs.(\ref{inf3}),
(\ref{rG}) and (\ref{rh}) we get
\begin{equation}
\rho_\gamma=\frac{R\,V'\,^2}{4\,\kappa\,V\,(1-V/\rho_c)\,(1+R)^2}
.\label{rh-1}
\end{equation}
Introducing the dimensionless slow-roll parameter, we write
\begin{equation}
\varepsilon\equiv-\frac{\dot{H}}{H^2}\simeq\,\frac{V'\,^2}{2\,\kappa\,(1+R)\,V^2}\,\left[\frac{(1-2V/\rho_c)}{(1-V/\rho_c)^2}\right]
,\label{ep}
\end{equation}
and the second slow-roll parameter  $\eta$ becomes
\begin{equation}
\eta\equiv-\frac{\ddot{H}}{H
\dot{H}}\simeq\,\frac{1}{\kappa\,V\,(1-V/\rho_c)\,(1+R)}\,\left[V''-\frac{2V'\,^2}{\rho_c\,(1-2V/\rho_c)}-
\frac{V'\,^2}{2V}\frac{(1-2V/\rho_c)}{(1-V/\rho_c)}\right].\label{eta}
\end{equation}

We see that for $R=0$ (or $\Gamma=0$), the parameters
$\varepsilon$ and $\eta$ given by Eqs.(\ref{ep}) and (\ref{eta})
respectively, are reduced  to the typical expression for cool
inflation in LQC\cite{good}. Note that the term in the bracket of
Eq.(\ref{ep}) is the correction to the standard warm inflationary
model.

It is possible to find a relation between the energy densities
$\rho_\gamma$ and $\rho_\phi$ given by
\begin{equation}
\rho_\gamma=\frac{R}{2(1+R)}\left[\frac{(1-\rho_\phi/\rho_c)}{(1-2\,\rho_\phi/\rho_c)}\right]\,\varepsilon\,\rho_\phi
\simeq\frac{R}{2(1+R)}\left[\frac{(1-V/\rho_c)}{(1-2\,V/\rho_c)}\right]\,\varepsilon\,V.\label{rho5}
\end{equation}
Recall that during inflation  the energy density of the scalar
field becomes dominated by the potential energy, i.e.
$\rho_\phi\sim V$.

The condition  which the warm inflation epoch on a LQC could take
place can be summarized with the parameter $\varepsilon$
satisfying  the inequality  $\varepsilon<1.$ This condition is
analogue to the requirement that  $\ddot{a}> 0$. The condition
given above is rewritten in terms of the densities by using
$\rho_\gamma$, we get
\begin{equation}
\left[\frac{(1-\rho_\phi/\rho_c)}{(1-2\,\rho_\phi/\rho_c)}\right]\,\rho_\phi>
\frac{2(1+R)}{R}\;\rho_\gamma.\label{cond}
\end{equation}

Inflation ends when the universe heats up at a time when
$\varepsilon\simeq 1$, which implies
\begin{equation}
\left[\frac{V_f\,'}{V_f}\right]^2\;\left[\frac{(1-2V_f/\rho_c)}{\kappa\,(1-V_f/\rho_c)^2}\right]\simeq\,2\,(1+R_f).\label{fin}
\end{equation}
The number of e-folds at the end of inflation is given by
\begin{equation}
N\simeq-\kappa\,\int_{\phi_{*}}^{\phi_f}\frac{V}{V'}\,(1-V/\rho_c)\,(1+R)
d\phi'.\label{N}
\end{equation}

In the following, the subscripts  $*$ and $f$ are used to denote
to the epoch when the cosmological scales exit the horizon and the
end of  inflation, respectively.

\section{Perturbations\label{sectpert}}

In this section we will study the scalar and tensor perturbations
for our model. Note that in the  case of scalar perturbations the
scalar and the radiation fields are interacting. Therefore,
isocurvature (or entropy) perturbations are generated besides the
adiabatic ones. This occurs because warm inflation can be
considered as an inflationary model with two basic fields
\cite{Jora1,Jora}. In this context dissipative effects  can
produce a variety of spectral, ranging between red and blue
\cite{62526,Jora}, and thus producing the running blue to red
spectral suggested by WMAP five-year data\cite{WMAP}.

As argued in Ref.\cite{good} for LQC, the density perturbation
could be written as
$\delta_H=\frac{2}{5}\frac{H}{\dot{\phi}}\,\delta\phi$
\cite{Liddle}. From Eqs.(\ref{inf3}) and (\ref{rG}), the latter
equation becomes
\begin{equation}
\delta_H^2=\frac{36}{25}\frac{H^4\,(1+R)^2}{V'\,^2}\,\delta\phi^2=
\frac{4}{25}\left(\frac{\kappa^2\,V^2\,(1-V/\rho_c)^2\,(1+R)^2}{V'\,^2}\right)\,\delta\phi^2.\label{331}
\end{equation}

The  scalar field presents  fluctuations which are due to the
interaction between the scalar and the radiation fields.  In the
case of strong  dissipation, the dissipation coefficient $\Gamma$
is much greater that the  rate expansion $H$ , i.e.
$R=\Gamma/3H\gg 1$ and following  Taylor and Berera\cite{Bere2},
we can write
\begin{equation}
(\delta\phi)^2\simeq\,\frac{k_F\,T_r\,}{2\,\pi^2},\label{del}
\end{equation}
where  the wave-number $k_F$ is defined by $k_F=\sqrt{\Gamma
H/V}=H\,\sqrt{3 \,R}\geq H$, and corresponds to the freeze-out
scale at which dissipation damps out to the thermally excited
fluctuations. The freeze-out wave-number $k_F$ is defined at the
point where the inequality $V_{,\,\phi\,\phi}< \Gamma H$, is
satisfied \cite{Bere2,Graham:2009bf}.

From Eqs. (\ref{331}) and (\ref{del}) it follows that
\begin{equation}
\delta^2_H\approx\;\frac{2}{25\,\pi^2}\,
\left[\frac{T_r}{V'\,^2}\right]\,\left[\kappa\,R\,V\,(1-V/\rho_c)\right]^{5/2}.\label{dd}
\end{equation}

The scalar spectral index $n_s$ is given by $ n_s -1 =\frac{d
\ln\,\delta^2_H}{d \ln k}$,  where the interval in wave number is
related to the number of e-folds by the relation $d \ln k(\phi)=-d
N(\phi)$. From Eq.(\ref{dd}), we get
\begin{equation}
n_s  \approx\,
1\,-\,\left[5\widetilde{\varepsilon}-2\widetilde{\eta}-\zeta
\right],\label{ns1}
\end{equation}
where, the  slow-roll parameters $\widetilde{\varepsilon}$,
$\widetilde{\eta}$ and $\zeta$, (for $R\gg 1$) are given by
\begin{equation}
\widetilde{\varepsilon}\approx\frac{1}{2\,\kappa\,R}\,
\left[\frac{V'}{V}\right]^2\;\frac{(1-2V/\rho_c)}{(1-V/\rho_c)^2},\label{e1}
\end{equation}
$$
\widetilde{\eta}\approx\frac{1}{\kappa\,V\,(1-V/\rho_c)\,R}\,\left[V''-\frac{2V'\,^2}{\rho_c\,(1-2V/\rho_c)}-
\frac{V'\,^2}{2V}\frac{(1-2V/\rho_c)}{(1-V/\rho_c)}\right],
$$
and
$$
\zeta\approx\frac{V'\,^2}{\kappa\,V\,(1-V/\rho_c)\,R}\,\left[\frac{4}{\rho_c(1-2V/\rho_c)}+\frac{(1-2V/\rho_c)}{V\,(1-V/\rho_c)}\right]
-\frac{5}{2}\frac{V'}{\kappa\,V\,(1-V/\rho_c)}\,\frac{R'}{R^2},
$$
respectively.

One of the interesting features of the five-year data set from
WMAP is that it hints at a significant running in the scalar
spectral index $dn_s/d\ln k=\alpha_s$ \cite{WMAP}.  From
Eq.(\ref{ns1}) we obtain that the running of the scalar spectral
index becomes

\begin{equation}
\alpha_s\approx\,2\,\widetilde{\varepsilon}\,\frac{V\,(1-V/\rho_c)}{V'\,(1-2V/\rho_c)}\,
\left[5\widetilde{\varepsilon}\,'-2\widetilde{\eta}\,'-\zeta\,'
\right].\label{dnsdk}
\end{equation}
In models with only scalar fluctuations the marginalized value for
the derivative of the spectral index is approximately $-0.05$ from
WMAP-five year data only \cite{WMAP}.

Tensor perturbation do not couple strongly to the thermal
background and so gravitational waves are only generated by
quantum fluctuations (as in standard inflation )\cite{Bere2}. The
corresponding spectrum  becomes
\begin{equation}
A^2_g=\,8\,\kappa\,\left(\frac{H}{2\pi}\right)^2=\frac{2\kappa^2\,}{3\,\pi^2}\,V\,(1-V/\rho_c).
\label{ag}
\end{equation}

For $R\gg1$ and from expressions (\ref{dd}) and (\ref{ag}) we may
write  the tensor-scalar ratio as
\begin{equation}
r(k)=\left.\left(\frac{A^2_g}{P_{\cal R}}\right)\right|_{\,k_*}
\simeq\left.\frac{4}{3\,\kappa^{1/2}}\,
\left[\frac{V'\,^2}{T_r\,V^{3/2}\,(1-V/\rho_c)^{3/2}\,R^{5/2}}\,\right]\right|_{\,k=k_*}.
\label{Rk}\end{equation} Here, $\delta_H\equiv\,2\,P_{\cal
R}^{1/2}/5$ and  $k_*$  is referred to $k=Ha$, the value when the
universe scale  crosses the Hubble horizon  during inflation.

Combining  WMAP observations \cite{WMAP} with the Sloan Digital
Sky Survey (SDSS) large scale structure surveys \cite{Teg}, it is
found an upper bound for $r$ given by $r(k_*\simeq$ 0.002
Mpc$^{-1}$)$ <0.28\, (95\% C.L.)$, where $k_*\simeq$0.002
Mpc$^{-1}$ corresponds to $l=\tau_0 k\simeq 30$, with the distance
to the decoupling surface $\tau_0$= 14,400 Mpc. The SDSS  measures
galaxy distributions at red-shifts $a\sim 0.1$ and probes $k$ in
the range 0.016 $h$ Mpc$^{-1}$$<k<$0.011 $h$ Mpc$^{-1}$. The
recent WMAP observation results give the values for the scalar
curvature spectrum $P_{\cal
R}(k_*)\equiv\,25\delta_H^2(k_*)/4\simeq 2.3\times\,10^{-9}$ and
the scalar-tensor ratio $r(k_*)<0.2$.
%We will make use
%of these values  to set constrains on the parameters for our
%model.

From Eqs.(\ref{dd}) and (\ref{e1}), we can write
\begin{equation}
V^{1/2}\,(1-\nu)^{1/2}\,(1-\nu)=\frac{4\,\pi^2\,P_{\cal
R}}{T_r\,(\kappa\,R)^{3/2}}\;\widetilde{\varepsilon},
\end{equation}
where $\nu=\nu(\phi)$ is defined by
$$
\nu=\frac{V(\phi)}{\rho_c}.
$$
Here, $\nu$ describe the quantum geometry effects in LQC, and is a
small quantity $\nu<10^{-9}$ (see Ref.\cite{good}). The
approximate value of the critical density $\rho_c$  in the
effective theory of LQC is $\rho_c\approx\,0.82\,\rho_p$, where
the Planck density $\rho_p=G^-2=m_p^4$, so we have
$\rho_c\approx\,0.82\,m_p^4$. By using the WMAP observations where
$P_{\cal R}\simeq 2.3\times\,10^{-9}$, and in view of $\nu\ll 1$,
we get
\begin{equation}
\nu\simeq\frac{16\,\pi^4\,P_{\cal
R}^2}{\rho_c\,T_r^2\,\kappa^3\,R^3}\,\widetilde{\varepsilon}\,^2\simeq
6.3\times
10^{-19}\,\frac{m_p^2}{T_r^2\,R^3}\,\widetilde{\varepsilon}\,^2.\label{c1}
\end{equation}
In the case of strong  dissipation $R\gg 1$, then we find from
Eq.(\ref{c1}) an upper limit for $\nu$, and it becomes
\begin{equation}
\nu\ll\,10^{-18}\,\left[\frac{m_p}{T_r}\right]^2\,\widetilde{\varepsilon}\,^2.\label{c2}
\end{equation}
Note that this inequality for $\nu$ become dependent of the
temperature of the thermal bath $T_r$ and
$\widetilde{\varepsilon}\,^2$. If we compared with respect to
standard supercooled inflation,  $\nu\simeq
10^{-9}\,\widetilde{\varepsilon}$ \cite{good}. Note also, that
this upper limit for $\nu$ increase when the temperature $T_r$
decreases.

\section{An example: Chaotic potential in the strong dissipation approach \label{exemple}}
Let us consider  an inflaton scalar field $\phi$  with a chaotic
potential. We write for the chaotic potential as $V=m^2\phi^2/2$,
where $m$ is the mass of the scalar field. An estimation of this
parameter is given for LQC  in Ref.\cite{good} and for warm
inflation in Ref.\cite{Bere2}. In the following, we develop the
model for a constant  dissipation coefficient
$\Gamma=\Gamma_0=const.$, and we will restrict ourselves to the
strong dissipation regime, i.e. $R\gg 1$.

By using the chaotic potential, we find that from Eq.(\ref{inf3})
\begin{equation}
\dot{\phi}=-\frac{m^2\,\phi}{\Gamma_0}\Longrightarrow\,\phi(t)=\phi_0\,e^{-m^2\,t/\Gamma_0},
\end{equation}
and  during the inflationary scenario the scalar field decays due
to dissipation into the radiation field. The Hubble parameter is
given by
\begin{equation}
H(t)=\frac{m\,\kappa^{1/2}\,\phi_0}{\sqrt{6}}\,e^{-m^2\,t/\Gamma_0}\,\left[1-\frac{m^2\,\phi_0^2}{2\,\rho_c}\,e^{-2m^2\,t/\Gamma_0}\right]^{1/2}
.
\end{equation}
Note that in the limit $\rho_c\gg\rho_\phi\simeq V$ the Hubble
parameter coincide with Ref.\cite{Bere2}. The dissipation
parameter $R$ in this case is
$$
R(t)=\frac{\sqrt{2}\,\Gamma_0}{\sqrt{3\,\kappa}\,m\,\phi_0}\,e^{m^2\,t/\Gamma_0}\,\left[1-\frac{m^2\,\phi_0^2}{2\,\rho_c}\,e^{-2m^2\,t/\Gamma_0}\right]^{-1/2}.
$$

By integrating Eq.(\ref{N}) the number of e-folds results in
\begin{equation}
N=-\frac{\Gamma_0\,\sqrt{\rho_c\,\kappa}}{2\sqrt{3}\,m^2}\,[h(\nu_f)-h(\nu_*)],\label{Nm}
\end{equation}
where $h(\nu)=\arcsin(\sqrt{\nu}\,)\,+\sqrt{\nu\,(1-\nu)}$.  From
the condition that $\widetilde{\varepsilon}\simeq 1$ (see
Eq.(\ref{fin})) at the end of warm-LQC inflation, we find that the
magnitude of $\nu$ at this time is
$\nu_f\approx\,\frac{3\,m^4}{\kappa\,\rho_c\,\Gamma_0^2}$.

By using Eq.(\ref{rho5}), we can relate the energy density of the
radiation field to the energy density of the inflaton field to
\begin{equation}
\rho_\gamma=\;\left(\frac{\sqrt{3}\,m^2}
{2\,\Gamma_0}\right)\,\left[\frac{\rho_\phi}{\kappa\,(1-\rho_\phi/\rho_c)}\right]^{1/2}.\label{9}
\end{equation}
Note again that in the limit $\rho_c\gg\rho_\phi$, Eq.(\ref{9})
coincides with that corresponding to the case where LQC is
absent\cite{Bere2}, i.e., $\rho_\gamma\propto\rho_\phi^{1/2}$.

From Eq.(\ref{dd}),  we obtain that the scalar power spectrum
becomes
\begin{equation}
P_{\cal R}(k)\approx\left.
\;\,\left(\frac{T_r}{4\,\pi^2\,m^2}\right)\,\left(\frac{\Gamma_0}{\sqrt{3}}\right)^{5/2}\;\left(\kappa^{5/4}\,V^{1/4}\,
\;\left[1-\nu\right]^{5/4}\right)\right|_{\,k=k_*},\label{ppp5}
\end{equation}
and from Eq.(\ref{Rk}) the tensor-scalar ratio is given by
\begin{equation}
r(k)\approx\;\left.\left(\frac{3^{1/4}\,8\,\kappa^{3/4}\,m^2}
{T_r\,\Gamma_0^{5/2}}\right)\;\left[\frac{V^3}{(1-\nu)}\right]^{1/4}\right|_{\,k=k_*}.\label{rrrr5}
\end{equation}

By using the WMAP observations where $P_{\cal R}(k_*)\simeq
2.3\times 10^{-9}$, $r(k_*)<0.2$ and $\nu\ll 1$, we obtained from
Eqs.(\ref{ppp5}) and (\ref{rrrr5}) that
\begin{equation}
\nu_*=\frac{V_*}{\rho_c}<\,4.2\times 10^{-12}.\label{V}
\end{equation}
From Eqs.(\ref{ppp5}) and (\ref{V}), we get the inequality
\begin{equation}
m^2<\,5\times
10^{4}\;\frac{T_r\,\Gamma_0^{5/2}}{m_p^{3/2}}.\label{A}
\end{equation}

Now we consider the special case in which we fix $T_r\simeq 0.24
\times 10^{16}$ GeV, and $\Gamma_0\simeq0.5\times 10^{13}$ GeV
(see Ref.\cite{Bhattacharya:2006dm}). In this special case we
obtained that the upper limit for the square mass of the scalar
field, is given by $m^2<5\times10^{-14}\,m_p^2$.

\begin{figure}[th]
\includegraphics[width=5.0in,angle=0,clip=true]{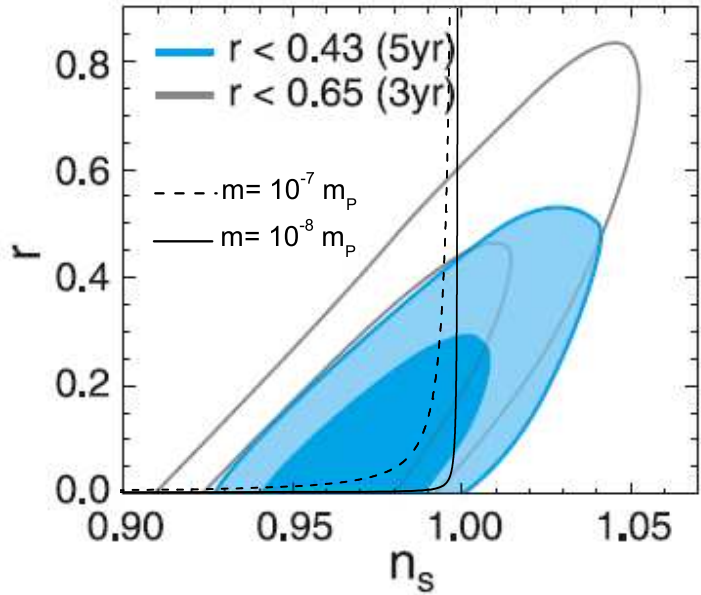}
\caption{ The plot shows $r$ versus $n_s$ for two values of $m$.
Here, we have fixed the values $T_r\simeq 0.24 \times 10^{16}$
GeV, $\Gamma_0\simeq0.5\times 10^{13}$ GeV, and $\rho_c\approx
0.82 m_p^4$, respectively. The five-year WMAP data places stronger
limits on the tensor-scalar ratio (shown in blue) than three-year
data (grey) \cite{Spergel}.  \label{fig1}}
\end{figure}

In Fig.(\ref{fig1}) we show the dependence of the tensor-scalar
ratio $r$ on the spectral index $n_s$, for the chaotic model
$V=m^2\phi^2/2$. From left to right $m=10^{-7}m_p$ (dashed  line)
and $m=10^{-8}m_p$ (solid line), respectively. From
Ref.\cite{WMAP}, two-dimensional marginalized
 constraints (68$\%$ and 95$\%$ confidence levels) on inflationary parameters
$r$ and $n_s$, the spectral index of fluctuations, defined at
$k_0$ = 0.002 Mpc$^{-1}$. The five-year WMAP data places stronger
limits on $r$ (shown in blue) than three-year data
(grey)\cite{Spergel}. In order to write down values that relate
$n_s$ and $r$, we used Eqs.(\ref{ns1}) and (\ref{rrrr5}).  Also we
have used  the values $T_r\simeq 0.24 \times 10^{16}$ GeV,
$\Gamma_0\simeq0.5\times 10^{13}$ GeV, and $\rho_c\approx 0.82
m_p^4$.

From Eqs.(\ref{Nm}) and (\ref{rrrr5}) , we observed  that for $m =
10^{-7}\,m_p$, the curve $r = r(n_s)$ (see Fig. (\ref{fig1})) for
WMAP 5-years enters the 95$\%$ confidence region  where the ratio
$r\simeq 0.42$, which corresponds to the number of e-folds, $N
\simeq 66.5$. For $m=10^{-8}\,m_p$, $r \simeq 0.44$ corresponds to
$N \simeq 694$.  From the 68$\%$ confidence region for $m=10^{-7}
m_p$ , $r\simeq 0.28$, which corresponds to $N\simeq$ 50.2. For
$m=10^{-8}\,m_p$, $r \simeq 0.29$ corresponds to $N \simeq 539$.

\section{Conclusions \label{conclu}}

In this paper we have investigated the  warm inflationary scenario
in LQC. In the slow-roll approximation we have found a general
relationship between the radiation and scalar field energy
densities. This has led us to a general criterium for
 warm inflation in LQC to occur (see Eq.(\ref{cond})).

Our  specific model is described by  a chaotic potential and we
have considered the case in which   the dissipation coefficient,
$\Gamma=\Gamma_0=$constant.  Here, we have
 found that the condition for $\nu_*$ presents the
same characteristic that  occurs in cool inflation for the LQC
\cite{good}, except that it depends on the extra parameter $T_r$.
In this case, we have obtained the  explicit expressions for the
corresponding scalar spectrum index and the running of the scalar
spectrum index. We also demonstrated that the scalar spectral
index, its running and the tensor-to-scalar ratio can be expressed
in terms of slow-roll parameters as well as the LQC parameter
$\nu$.

In order to bring some explicit results we have taken the
constraint in the $r -n_s$   plane to the chaotic model,
$V=m^2\phi^2/2$. We  noted that the parameter $m$, which is
bounded from below, $m^2<\,5\times 10^{-14}\;m_p^2$, (see
Eq.(\ref{A})) and  the model is well supported by the data as
could be seen from Fig.(1). Here, we have used  the values
$T_r\simeq 0.24 \times 10^{16}$ GeV, $\Gamma_0\simeq0.5\times
10^{13}$ GeV, and $\rho_c\approx 0.82 m_p^4$, respectively. On the
other hand, by using the WMAP observations where $P_{\cal
R}(k_*)\simeq 2.3\times 10^{-9}$, $r(k_*)<0.2$ and $\nu\ll 1$, we
obtained from Eqs.(\ref{ppp5}) and (\ref{rrrr5}) that
$\nu_*<\,4\times 10^{-12}$. We should note that this inequality
for $\nu_*$, becomes small by three order of magnitude when it is
compared with the case of standard-LQC\cite{good}.

We should  note that other properties of this model deserve
further study. For example,  we have not addressed the
non-Gaussian effects during warm inflation (see e.g.,
Refs.\cite{vanishing,19,25}). A possible calculation from the
non-linearity parameter $f_{NL}$, would give new constrains on the
parameters of the model. Also, a sophisticated analysis  would
give new constraints on the dissipative coefficient
$\Gamma=\Gamma(\phi,T_r)$,  the cosmological perturbations in
LQC\cite{per}, and for warm inflation, see, e.g.,\cite{ramon}. We
hope to return to this point in the near future.

\begin{acknowledgments}
 This work  was supported by COMISION NACIONAL DE CIENCIAS Y TECNOLOGIA
through FONDECYT Grant N$^{0}$. 1090613,  also from  PUCV DI-PUCV
2009.
\end{acknowledgments}

%\\\\\\\\\\\\\\\\\\\\\\\\\\\\\\\\\\\\\\\\\\\\\\\\\\\\\\\\\\\\\\\\\\\\\\\


\begin{thebibliography}{99}



%%%%%Refer introduction%%%%%%%%%%%%%%%%%%%%%%%%%%%%%%%%%%





\bibitem{warm}
A. Berera,   Phys. Rev. Lett. {\bf 75}, 3218 (1995); A. Berera,
   Phys. Rev. D {\bf 55}, 3346 (1997).




\bibitem{62526}
L.M.H. Hall, I.G. Moss  and A. Berera,   Phys.Rev.D {\bf 69},
083525 (2004); I.G. Moss,  Phys.Lett.B {\bf 154}, 120 (1985); A.
Berera  and L.Z. Fang, Phys.Rev.Lett. {\bf 74} 1912 (1995); A.
Berera,  Nucl.Phys B {\bf 585}, 666 (2000).

\bibitem{1126}A. Berera,   Phys. Rev.D {\bf 54},
2519 (1996).

\bibitem{Berera:2008ar}
  A.~Berera, I.~G.~Moss and R.~O.~Ramos,
  %``Warm Inflation and its Microphysical Basis,''
  Rept.\ Prog.\ Phys.\  {\bf 72}, 026901 (2009).

\bibitem{jj} M.~Bastero-Gil and A.~Berera,
  %``Warm inflation model building,''
  Int.\ J.\ Mod.\ Phys.\  A {\bf 24}, 2207 (2009).


\bibitem{taylorberera} A.~Berera, M.~Gleiser and R.~O.~Ramos,
  %``A First Principles Warm Inflation Model that Solves the Cosmological
  %Horizon/Flatness Problems,''
  Phys.\ Rev.\ Lett.\  {\bf 83}, 264 (1999);   J. Mimoso, A. Nunes  and D. Pavon,  Phys.Rev.D {\bf
73}, 023502 (2006); S.~del Campo and R.~Herrera,
  %``Warm-Chaplygin inflationary universe model,''
  Phys.\ Lett.\  B {\bf 665}, 100 (2008);
R.~Herrera, S.~del Campo and J.~Saavedra,
  %``Tachyon Warm Inflationary Universe Model In The Weak Dissipation Regimen,''
  J.\ Phys.\ Conf.\ Ser.\  {\bf 134}, 012008 (2008);  S.~del Campo, R.~Herrera and J.~Saavedra,
  %``Tachyon warm inflationary universe model in the weak dissipative regime,''
  Eur.\ Phys.\ J.\  C {\bf 59}, 913 (2009).





%%%%%%%%%%%%%%%%%%%%%%%%%%%%%%%%DGP%%%%%%%%%%%%%%%%%%%%




%%%%%%%%%%%%%%%%%%%%%%%%%%%%%%%%%%%%%%%%%%






%%%%%%%%%%%%%%%%%%%%%%%%%%%%%%%%%%%%%%%Refere Sec II
\bibitem{5} T. Thiemann, Lect. Notes Phys. {\bf631}, 41 (2003); A. Ashtekar and
J. Lewandowski, Class. Quant. Grav. {\bf21}, R53 (2004); C.
Rovelli, Quantum Gravity, Cambridge Monographs on Mathematical
Physics (2004).

\bibitem{6} M. Bojowald, Living Rev. Rel. {\bf8}, 11 (2005).

\bibitem{7} M. Bojowlad, Phys. Rev. Lett. {\bf86}, 5227 (2001).

\bibitem{8} A. Ashtekar, M. Bojowald, J. Lewandowski, Adv. Theor. Math.
Phys. {\bf7}, 233 (2003).

\bibitem{9} M. Bojowald, G. Date, K. Vandersloot, Class. Quantum Grav. {\bf21},
1253 (2004); P. Singh, A. Toporensky, Phys. Rev. D {\bf69}, 104008
(2004); G.V. Vereshchagin, JCAP {\bf07}, 013(2004); G. Date, Phys.
Rev. D {\bf71}, 127502 (2005). M. Bojowald, Phys. Rev. Lett.
{\bf95}, 061301 (2005); M. Bojowald, R. Goswami, R. Maartens, P.
Singh, Phys. Rev. Lett. {\bf95}, 091302 (2005); G. Date, Phys.
Rev. D {\bf71}, 127502 (2005); G. Date, G. M. Hossain, Phys. Rev.
Lett. {\bf94}, 011302 (2005); A. Ashtekar, M. Bojowald,
Class.Quant.Grav. {\bf23}, 391 (2006); R. Goswami, P. S. Joshi, P.
Singh, Phys. Rev. Lett. {\bf96}, 031302 (2006).



\bibitem{AA} A. Ashtekar, T. Pawlowski, P. Singh and K. Vandersloot, Phys. Rev. D {\bf75}, 024035
(2007); L. Szulc ,W. Kaminski  and J. Lewandowski, Class. Quant.
Grav. {\bf24}, 2621 (2007); K. Vandersloot,  Phys. Rev. D {\bf75},
023523 (2007); L. Szulc,
 Class. Quant. Grav.
{\bf24}, 6191 (2007).




\bibitem{good} X.~Zhang and Y.~Ling,
  %``Inflationary universe in loop quantum cosmology,''
  JCAP {\bf 0708}, 012 (2007).



\bibitem{int1}S.~Chen, B.~Wang and J.~Jing,
  %``Dynamics of interacting dark energy model in Einstein and Loop Quantum
  %Cosmology,''
  Phys.\ Rev.\  D {\bf 78}, 123503 (2008).


\bibitem{int2} P.~Wu and S.~N.~Zhang,
  %``Cosmological evolution of interacting phantom (quintessence) model in Loop
  %Quantum Gravity,''
  JCAP {\bf 0806}, 007 (2008).

\bibitem{WMAP} J.~Dunkley {\it et al.},
  %``Five-Year Wilkinson Microwave Anisotropy Probe (WMAP) Observations:
  %Likelihoods and Parameters from the WMAP data,''
  Astrophys.\ J.\ Suppl.\  {\bf 180}, 306 (2009);  G.~Hinshaw {\it et al.},
  %``Five-Year Wilkinson Microwave Anisotropy Probe (WMAP) Observations: Data
  %Processing, Sky Maps, and Basic Results,''
  Astrophys.\ J.\ Suppl.\  {\bf 180}, 225 (2009).

\bibitem{super} M.~Bojowald,
  %``Isotropic loop quantum cosmology,''
  Class.\ Quant.\ Grav.\  {\bf 19}, 2717 (2002);
M.~Bojowald,
  %``Inflation from Quantum Geometry,''
  Phys.\ Rev.\ Lett.\  {\bf 89}, 261301 (2002); S.~Tsujikawa, P.~Singh and R.~Maartens,
  %``Loop quantum gravity effects on inflation and the CMB,''
  Class.\ Quant.\ Grav.\  {\bf 21}, 5767 (2004);
 P.~Singh, K.~Vandersloot and G.~V.~Vereshchagin,
  %``Non-singular bouncing universes in loop quantum cosmology,''
  Phys.\ Rev.\  D {\bf 74}, 043510 (2006); G.~Calcagni and M.~Cortes,
  %``Inflationary scalar spectrum in loop quantum cosmology,''
  Class.\ Quant.\ Grav.\  {\bf 24}, 829 (2007); A.~Ashtekar,
  %``An Introduction to Loop Quantum Gravity Through Cosmology,''
  Nuovo Cimento Soc. Ital. Fis. {\bf B\, 122}, 135 (2007).










\bibitem{38}P. Singh, Phys. Rev. D {\bf 73}, 063508 (2006).
\bibitem{44}M. Sami, P. Singh and S. Tsujikawa, Phys. Rev. D {\bf74}, 043514 (2006)

\bibitem{52}K. Vandersloot, Phys. Rev. D {\bf71}, 103506 (2005);
P. Singh and K. Vandersloot, Phys. Rev. D {\bf72}, 084004 (2005).

\bibitem{Bh}A.~Ashtekar, J.~Baez, A.~Corichi and K.~Krasnov,
  %``Quantum geometry and black hole entropy,''
  Phys.\ Rev.\ Lett.\  {\bf 80}, 904 (1998)
\bibitem{Moss}I. G. Moss and C. Xiong, arXiv:hep-ph/0603266.



\bibitem{Zhang:2009ge}
  Y.~Zhang,
  %``Warm Inflation with a General Form of the Dissipative Coefficient,''
  JCAP {\bf 0903}, 023 (2009).
\bibitem{delCampo:2009xi}
  S.~del Campo and R.~Herrera,
  %``Warm-Intermediate inflationary universe model,''
  JCAP {\bf 0904}, 005 (2009).


%%%%%%%%%%%%%%%%%%%%%%%%%%Perurbation escalares
\bibitem{new0} U. Weiss, Quantum Dissipative Systems (World Scientific, Singapore,
1993).


\bibitem{new} A. Berera and R. Ramos, Phys. Lett. B {\bf 567}, 294
(2003); A. Berera and R. Ramos, Phys.Rev. D {\bf 71}, 023513
(2005); M. Bastero-Gil and A. Berera, Phys.Rev. D {\bf 72}, 103526
(2005).

\bibitem{new2} I.~G.~Moss and C.~Xiong,
  %``Non-gaussianity in fluctuations from warm inflation,''
  JCAP {\bf 0704}, 007 (2007).


\bibitem{Jora} A. Starobinsky  and J. Yokoyama, Density fluctuations
in Brans-Dicke inflation, Published in the Proceedings of the
Fourth Workshop on General Relativity and Gravitation. Edited by
K. Nakao, et al. Kyoto, Kyoto University, 1995. pp. 381,
gr-qc/9502002; A. Starobinsky  and S. Tsujikawa, Nucl.Phys.B {\bf
610}, 383 (2001); H. Oliveira  and S. Joras,    Phys. Rev. D {\bf
64}, 063513 (2001).

\bibitem{Jora1} H. Oliveira,   Phys. Lett. B {\bf 526}, 1 (2002).








\bibitem{Liddle} A. Liddle  and D. Lyth, Cosmological inflation and
large-scale structure, 2000, Cambridge University;J. Linsey ,
 A. Liddle,  E. Kolb and E. Copeland,    Rev. Mod. Phys {\bf 69}, 373 (1997);
B. Bassett, S. Tsujikawa  and D. Wands,   Rev. Mod. Phys. {\bf
78}, 537 (2006).

\bibitem{Bere2} A. Taylor  and A. Berera,   Phys. Rev. D {\bf 62},
083517 (2000).

\bibitem{Graham:2009bf}
  C.~Graham and I.~G.~Moss,
  %``Density fluctuations from warm inflation,''
  JCAP {\bf 0907}, 013 (2009)


%%%%%%%%%%%%%%%%%%%%%Pert. Waves
%\bibitem{Bha}K. Bhattacharya, S. Mohanty  and A. Nautiyal,
% Phys.Rev.Lett. {\bf 97}, 251301 (2006).


\bibitem{Teg}M. Tegmark  et al.,  Phys. Rev. D {\bf 69},
103501 (2004).
\bibitem{Bhattacharya:2006dm}
  K.~Bhattacharya, S.~Mohanty and A.~Nautiyal,
  %``Enhanced polarization of CMB from thermal gravitational waves,''
  Phys.\ Rev.\ Lett.\  {\bf 97}, 251301 (2006).


\bibitem{Spergel} D.~N.~Spergel {\it et al.}, Astrophys.\ J.\
Suppl.\  {\bf 170}, 377 (2007).

%%%%%%%%%%%%%%%%%%%%%Aplicaciones
%\bibitem{yo}R. Herrera, S. del Campo  and C. Campuzano, JCAP 0610, 009 (2006).

%\bibitem{yo1}S. del Campo,R. Herrera  and D. Pavon,
%  Phys. Rev. D {\bf 75}, 083518 (2007).






\bibitem{vanishing} A. Gangui, F. Lucchin, S. Matarrese  and
S. Mollerach,   Astrophys. J. \textbf{430}, 447 (1994); S. Gupta,
Phys. Rev. D \textbf{73}, 083514 (2006);  I. Moss  and  C. Xiong,
JCAP 0704, 007 (2007).


\bibitem{19}J. C. Bueno Sanchez, M. Bastero-Gil, A. Berera and K. Dimopoulos,
Phys. Rev. D {\bf77}, 123527 (2008).

\bibitem{25}S. Gupta, A. Berera, A. F. Heavens and S. Matarrese, Phys. Rev. D \textbf{66}, 043510 (2002);
I. G. Moss and C. Xiong, JCAP \textbf{0704}, 007 (2007).

\bibitem{per} M.~Bojowald, H.~H.~Hernandez, M.~Kagan, P.~Singh and A.~Skirzewski,
  %``Hamiltonian cosmological perturbation theory with loop quantum gravity
  %corrections,''
  Phys.\ Rev.\  D {\bf 74}, 123512 (2006); M.~Bojowald, H.~Hernandez, M.~Kagan, P.~Singh and A.~Skirzewski,
  %``Formation and evolution of structure in loop cosmology,''
  Phys.\ Rev.\ Lett.\  {\bf 98}, 031301 (2007); M.~Bojowald, G.~M.~Hossain, M.~Kagan and S.~Shankaranarayanan,
  %``Gauge invariant cosmological perturbation equations with corrections from
  %loop quantum gravity,''
  Phys.\ Rev.\  D {\bf 79}, 043505 (2009).



\bibitem{ramon} R.~Herrera, S.~del Campo and C.~Campuzano,
  %``Tachyon warm inflationary universe models,''
  JCAP {\bf 0610}, 009 (2006); S.~del Campo, R.~Herrera and D.~Pavon,
  %``Cosmological perturbations in warm inflationary models with viscous
  %pressure,''
  Phys.\ Rev.\  D {\bf 75}, 083518 (2007);  M.~A.~Cid, S.~del Campo and R.~Herrera,
  %``Warm inflation on the brane,''
  JCAP {\bf 0710}, 005 (2007); S.~del Campo and R.~Herrera,
  %``Warm inflation in the DGP brane-world model,''
  Phys.\ Lett.\  B {\bf 653}, 122 (2007).





\end{thebibliography}
\end{document}